\documentclass[aps,prb,twocolumn,showpacs,floatfix]{revtex4}
\usepackage{amsmath,amssymb,graphicx}
\usepackage{epsfig}
\usepackage{graphicx}
\usepackage{enumerate}
\usepackage[toc,page,header]{appendix}

\newcommand{\beq}{\begin{equation}}
\newcommand{\eeq}{\end{equation}}

\newcommand{\bea}{\begin{eqnarray}}
\newcommand{\eea}{\end{eqnarray}}

\begin{document}

%\title{Phase rigidity in double dot Aharonov-Bohm interferometer coupled to dephasing and voltage probes}

%\title{Symmetries of current in %nonequilibrium
%double-dot Aharonov-Bohm interferometers with dephasing and inelastic effects
%including a voltage probe
%}
\title{
Magnetic field symmetries of nonlinear transport with elastic and inelastic
scattering
}
% Relations between deviations from Onsager
%symmetry and dc-rectification in Aharonov-Bohm interferometers}
\author{Salil Bedkihal$^1$, Malay Bandyopadhyay$^2$, Dvira Segal$^1$ }
\affiliation{$^1$Chemical Physics Theory Group, Department of
Chemistry, University of Toronto, 80 Saint George St. Toronto,
Ontario, Canada M5S 3H6}
\address{$^2$
School of Basic Sciences, Indian Institute of Technology Bhubaneswar, 751007, India}

%73.23.-b,
% Electronic transport in mesoscopic systems
%73.40.-c,
%Electronic transport in interface structures
%73.63.Nm,
%Quantum wires
%73.63.-b,
%Electronic transport in nanoscale materials and structures
%85.65.+h,
%Molecular electronic devices
% 72.10.-d  Theory of electronic transport; scattering mechanisms
% 73.50.Fq  High-field and nonlinear effects
\begin{abstract}
We study nonlinear electronic transport symmetries in Aharonov-Bohm interferometers subjected to inelastic scattering effects and show that odd (even) conductance terms are even (odd) in the magnetic field when the junction is (left-right) spatially symmetric. This observation does not hold when an asymmetry is introduced,
as we show numerically, but odd conductance terms only manifest a weak breakdown of the
magnetic field symmetry.
%However, while spatial asymmetry leads to a strong breakdown of the even conductance terms symmetry, odd conductance terms show only a ``weak" symmetry violation.
Under elastic dephasing effects, the Onsager-Casimir symmetry is maintained beyond linear
response and under spatial asymmetries.

%We prove an exact relation between nonlinear transport contributions and
%deviations from phase rigidity in Aharonov-Bohm interferometers subjected to
% inelastic scattering effects.
%We consider a geomtrically symmetric two-terminal Aharonov-Bohm ring structure
%coupled to a third terminal, serving as a voltage probe,
%phenomenologically introducing inelastic effects, and prove that the current
%rectification {\it equals} the deviation from phase rigidity.
%In other words, even conductance terms are proved to non-even in the magnetic flux.
%In contrast, in asymmetric junctions we show, using exact numerical simulations,
%that the non-even terms (with respect to the magnetic field) are not necesserally even in the  voltage drop.
\end{abstract}

\date{\today}

\maketitle

%======================================================================

{\it Introduction.}
The Onsager-Casimir symmetry relations \cite{OnsagerC}
hold close to equilibrium, implying that the two-probe linear
conductance is an even function of the magnetic field $B$.
As a consequence, the two-terminal transmission function of coherent conductors
satisfies $\mathcal T(B)=\mathcal T(-B)$.
Within Aharonov-Bohm (AB) interferometers,
this symmetry is displayed  by the ``phase rigidity"
of the conductance oscillations with $B$. \cite{Imry,Yacoby}
Beyond linear response,
the phase symmetry of the conductance, or reciprocity theorem,
is generally not enforced, and several experimental works
\cite{Linke,breakE1,breakE2,breakE3,breakE4} have demonstrated its breakdown.
Supporting theoretical studies
have incorporated many-body interactions \cite{break1, break2, break3, Kubo, Meir}, %of the system with either internal or external degrees of freedom,
but typically approached the problem
by calculating the screening potential within the
conductor self consistently, a procedure
often limited to low-order conduction terms \cite{break1, break3, Kubo}.

In this paper we aim in generalizing the reciprocal relations to
the nonlinear transport regime, while allowing for inelastic scattering effects.
Phase-breaking and energy dissipation processes arise due to the
interaction of electrons with other degrees of freedom, electrons,
phonons, and defects. Here we incorporate such processes
phenomenologically, by using the well-established and experimentally feasible method of
B\"uttiker dephasing and voltage probes \cite{Buttiker,Vexp}.
For spatially symmetric junctions we then
discuss the exact symmetry relations
beyond the Onsager symmetry, and their violation, addressing all transport coefficients at the same footing.
%
%In this work we introduce phase breaking and energy dissipation processes
%into the electron dynamics using the well-established method of
%B\"uttiker voltage probe \cite{XXX}.
%In this technique, inelastic effects are emulated by
%including a third (probe) terminal,
%demanding that the
%total-net charge current to this probe vanishes.
%A stronger condition, enforcing the charge curent in the probe to vanish
%at a given electron energy, allows for dephasing processes only.
%
Expanding the current $I(\phi)$ in powers of the bias $\Delta \mu$, we write
\bea
I(\phi)=G_1(\phi) \Delta\mu + G_2(\phi) (\Delta\mu)^2  + G_3(\Delta\mu)^3 +...
\label{eq:III}
\eea
with $G_{n>1}$ as the nonlinear conductance coefficients.
Here we have introduced the AB phase $\phi=2\pi\Phi/\Phi_0$, $\Phi$ is the magnetic flux
threading through the AB ring and $\Phi_0=h/e$ is the magnetic flux
quantum. In this work we study relations between two quantities: a measure for
the magnetic field asymmetry
\bea
\Delta I(\phi)\equiv [I(\phi)-I(-\phi)]/2,
\eea
and the dc-rectification current,
\bea
\mathcal R(\phi)\equiv \frac{1}{2}[I(\phi)+\bar I(\phi)] = G_2(\phi) (\Delta\mu)^2  + G_4(\phi) (\Delta\mu)^4+ ...
\eea
with $\bar I$ defined as the current obtained upon interchanging
the chemical potentials of the two terminals (assuming identical temperatures).
We also study the behavior of odd conductance terms,
%
%\bea
$\mathcal D(\phi)\equiv  G_1(\phi) \Delta\mu  + G_3(\phi) (\Delta\mu)^3+ ...$
%\eea
%
%
%===================
\begin{figure}[b]
\vspace{-6mm}\hspace{2mm}
{ \hbox{\epsfxsize=40mm \epsffile{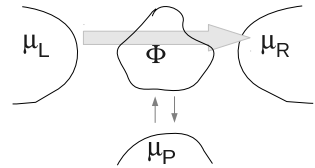}}}
\vspace{-1mm}
\caption{Scheme of our setup. The horizontal arrow represents the
charge current $I$. The two parallel arrows represent  (same magnitude) currents
into and from the $P$
terminal, serving to induce elastic and or inelastic scattering effects.
}
\label{FigS}
\end{figure}
%
%----------------------------------
For a non-interacting and a spatially symmetric system we expect
%
%\bea
$I(\phi)=-\bar I(-\phi)$
%\eea
%
to hold.
Combined with Eq. (\ref{eq:III}) we immediately note that
$G_{2n+1}(\phi)=G_{2n+1}(-\phi)$ and $G_{2n}(\phi)=-G_{2n}(-\phi)$ with $n$ as an integer.
We show below that these relations are obeyed in a symmetric junction even when many-body interactions
(inelastic scattering),  which depend on the applied bias in a nonlinear manner, are included.
%scattering events are introduced.

The principal results of this paper are now listed:
(i) Under elastic dephasing effects  we
prove that $\Delta I=0$ and $\mathcal R=0$, thus the current displays an even symmetry with respect to the magnetic field.  % as expected \cite{Buttiker-4}.
We then incorporate inelastic effects and show three results:  (ii) In the
absence of magnetic flux or for $\phi=2\pi k$, $k$ is an integer, no
rectification takes place, $\mathcal R(\phi=2\pi k)=0$. (iii) With
magnetic flux, in geometrically symmetric junctions, $\Delta
I(\phi)=\mathcal R(\phi)=-\mathcal R(-\phi)$.
%In other words, odd (even) conductance coefficients are even (odd) in the magnetic flux.
%This is not trivial since in our formalism inelastic scattering effects are effectively bias dependent.
%This observation corroborates with recent experiments of transport in
%likely geometrically asymemtric AB ring structures  \cite{breakE2}.
(iv) Using a double-dot interferometer, we demonstrate that
under a spatial asymmetry both even and odd conductance
terms show no particular magnetic field symmetry, however,
odd conductance terms are close to be symmetric with the magnetic field.
These observations corroborate with  transport experiments on
AB structures with
apparently a geometrical asymmetry in the ring-lead coupling  \cite{breakE2}.
Note that ``spatial" or ``geometrical" symmetry refers here to the left-right
symmetry of the junction. For simplicity, we set $e$=1, $h=1$
%$\hbar=1$
$k_B=1$, and ignore electron spin.
%=======================================

{\it Basic Expressions.}
Elastic dephasing effects and inelastic scattering processes are
implemented here using
the dephasing probe and the voltage probe techniques, respectively
 \cite{Buttiker}.
Particularly, we consider a setup including three terminals, $L$, $R$ and $P$,
with the $P$ terminal serving as the probe, see Fig. \ref{FigS}.
Our analysis relies on two exact relations:
The transmission coefficient from the $\xi$ to the
$\nu$ reservoir obeys time reversal symmetry,
\bea
\mathcal T_{\xi,\nu}(\epsilon,\phi)= \mathcal T_{\nu,\xi}(\epsilon,-\phi),
\label{eq:R1}
\eea
and the total probability is conserved ($\xi,\nu=L,R,P$) \cite{Kubo},
\bea
\sum_{\xi\neq \nu} \mathcal T_{\xi,\nu}(\epsilon,\phi)=
\sum_{\xi\neq \nu} \mathcal T_{\nu,\xi}(\epsilon,\phi).
\label{eq:R2}
\eea
%
%The proof of this second relation is included in e.g., Ref.  g\cite{Kubo}.
We focus on the steady-state tunneling current from the
$L$ reservoir into the system, $I(\phi)=I_L(\phi)$ with
\bea
I_L(\phi)&=&
\int_{-\infty}^{\infty} d\epsilon \Big[ \mathcal T_{L,R}(\epsilon,\phi)f_L(\epsilon)-
 \mathcal T_{R,L}(\epsilon,\phi)f_R(\epsilon)
\nonumber\\
&+& \mathcal T_{L,P}(\epsilon,\phi)f_L(\epsilon)-
 \mathcal T_{P,L}(\epsilon,\phi)f_P(\epsilon,\phi)
\Big],
\label{eq:IL}
\eea
written here assuming the Landauer's picture of noninteracting electrons.
%The current is expressed by the
%transmission function $\mathcal T_{\nu,\xi}(\epsilon)$,  given
%in terms of the Green's function of the system and the self energy matrices,
%see an example below.
The Fermi-Dirac distribution function
$f_{\nu}(\epsilon)=[e^{\beta_{\nu}(\epsilon-\mu_{\nu})}+1]^{-1}$ is
defined in terms of the chemical potential $\mu_{\nu}$ and the
inverse temperature $\beta_{\nu}$. In what follows we assume that the
temperature is identical in all terminals.
The current from the probe terminal to the system is given by %\cite{comm}
\bea &&I_P(\phi)= \int_{-\infty}^{\infty}d\epsilon \Big[\mathcal
T_{P,L}(\epsilon,\phi)f_P(\epsilon,\phi) - \mathcal
T_{L,P}(\epsilon,\phi)f_L(\epsilon) \Big]
\nonumber\\
&&+\int_{-\infty}^{\infty}d\epsilon \Big[ \mathcal
T_{P,R}(\epsilon,\phi)f_P(\epsilon,\phi) - \mathcal
T_{R,P}(\epsilon,\phi)f_R(\epsilon) \Big].
\label{eq:IP} \eea
The probe distribution function, generally phase dependent,
is determined by the probe condition, as we explain below.
For convenience, we simplify next our notation
by dropping the reference to the energy of the incoming electron
$\epsilon$ from both transmission functions and distribution
functions, and not putting limits of integrations which
are all evaluated between $\pm \infty$.
Finally, we do not explicitly include $\phi$ in $\mathcal T_{\xi,\nu}$ and $f_P$,
all evaluated at $\phi$, unless otherwise mentioned \cite{comm}.
%When we need to consider
%the transmission function $\mathcal T_{\nu,\xi}(-\phi)$, we usually
%put instead
%the complementary representation, $\mathcal T_{\xi,\nu}(\phi)$.
Using Eq. (\ref{eq:IL}), we identify the deviation from
the magnetic-field symmetry as
\bea
\Delta I&=&
\int \frac{1}{2}\left[ \mathcal T_{L,R} - \mathcal T_{R,L}\right] f_Rd\epsilon
\nonumber\\
&+&
\int \frac{1}{2}\left[ \mathcal T_{L,P}f_P(-\phi) - \mathcal T_{P,L} f_P(\phi)
\right]d\epsilon.
\label{eq:DeltaI}
\eea
The rectification contribution is written as
\bea \mathcal R &=& \int \frac{\mathcal T_{P,L}-\mathcal
T_{P,R}}{4}(f_L+f_R-f_P-\bar f_P) d\epsilon\label{eq:R} \eea
with $\bar f_P$ as the probe distribution when the biases $\mu_L$ and
$\mu_R$ are interchanged.

{\it (i) Elastic dephasing effects.}
%$\mathcal R=0$ and $\Delta I=0$.}
We implement elastic dephasing effects by demanding that the
energy-resolved particle current in the probe diminishes,
$I_P(\epsilon)=0$, with $I_P=\int I_P(\epsilon)d\epsilon$. Eq.
(\ref{eq:IP}) then provides the corresponding probe distribution
\bea
f_P(\phi)=\frac{\mathcal T_{L,P}f_L +  \mathcal T_{R,P}f_R   }{
\mathcal T_{P,L} + \mathcal T_{P,R}}.
\label{eq:dephP}
\eea
As highlighted, this function depends on the magnetic flux.
It is not difficult to prove that Onsager symmetry is satisfied here, beyond linear response. We
plug $f_P$ into Eq.  (\ref{eq:DeltaI}) and find,
%
%\bea \Delta I&=& \frac{1}{2}\int  \left[ \mathcal T_{L,R} - \mathcal
%T_{R,L}\right] f_R d\epsilon
%\nonumber\\
%&+&  \frac{1}{2}
%\int d\epsilon
%\mathcal T_{L,P}
%\frac{ \mathcal T_{P,L} f_L + \mathcal T_{P,R}f_R }{ \mathcal T_{L,P} +
%\mathcal T_{R,P}}
%\nonumber\\
%&-&\frac{1}{2}
%\int  d\epsilon
%\mathcal T_{P,L}
%\frac{ \mathcal T_{L,P} f_L + \mathcal T_{R,P}f_R }{ \mathcal T_{P,L} +
%\mathcal T_{P,R}} %\Big]
%\nonumber
%\eea
%
%Since the denominators are identical, see Eq. (\ref{eq:R2}), we
%combine the last two terms and get %%
after simple algebraic manipulations,
\bea
\Delta I&=& \frac{1}{2}
\int  \left[ \mathcal T_{L,R} - \mathcal T_{R,L}\right] f_R d\epsilon
\nonumber\\
&+&\frac{1}{2}
\int
\frac{ \left[ \mathcal T_{L,P} \mathcal T_{P,R} - \mathcal T_{P,L} \mathcal T_{R,P} \right]f_R  }
{ \mathcal T_{P,R} +
\mathcal T_{P,L}} d\epsilon.
\eea
Utilizing Eq. (\ref{eq:R2}) in the form
$\mathcal T_{L,P}$ = $\mathcal T_{P,L} + \mathcal T_{P,R}-\mathcal T_{R,P}$,
we organize the numerator of the second integral,
$(\mathcal T_{P,R}  - \mathcal T_{R,P} )   (\mathcal T_{P,R}  + \mathcal T_{P,L} ) f_R $.
This results in
\bea
\Delta I&=&
\frac{1}{2}\int  \left[ \mathcal T_{L,R} - \mathcal T_{R,L}
+
\mathcal T_{P,R}  - \mathcal T_{R,P} \right] f_R d\epsilon
%\nonumber\\
%&=&
%\int  f_R \left[\sum_{\nu\neq R}\mathcal T_{\nu,R} -
%\sum_{\nu\neq R}\mathcal T_{R,\nu}\right] d\epsilon,
\eea
which is identically zero, given Eq. (\ref{eq:R2}). This concludes
our proof that dephasing effects (implemented via a dephasing probe)
cannot break the reciprocity theorem even in the nonlinear regime. Following
similar steps we can show that $\mathcal R=0$ under dephasing effects:
We substitute $f_P$ into Eq. (\ref{eq:IL}) and obtain
%
%\bea
$I_L=\int[ F_L f_L - F_R f_R] d\epsilon$
%\eea
%
with $F_L=[\mathcal T_{L,R}(\mathcal T_{P,L}+\mathcal T_{P,R})
+ \mathcal T_{L,P}\mathcal T_{P,R}]/(\mathcal T_{P,L}+\mathcal T_{P,R})$.
$F_R$ is defined analogously, interchanging  $L$ by $R$.
Using Eq. (\ref{eq:R2}), one can show that $F_L=F_R$, thus $I=-\bar I$
and $\mathcal R=0$.
To conclude, $I= \mathcal D(\phi)=\mathcal D(-\phi)$ under elastic dephasing
and  spatial asymmetry.

{\it Inelastic effects.} We introduce inelastic effects using the
voltage probe technique, where we demand that the net-total particle
current flowing in the $P$ reservoir is zero, $I_P=0$. This choice
allows for energy exchange processes within the $P$ reservoir. The
probe condition  produces the following relations
\bea &&\int d\epsilon (\mathcal T_{P,L}+\mathcal T_{P,R})f_ P(\phi)
= \int d\epsilon (\mathcal T_{L,P}f_L +\mathcal T_{R,P}f_R)
\nonumber\\
&&\int d\epsilon  (\mathcal T_{L,P}+\mathcal T_{R,P}) f_P(-\phi) =
\int d\epsilon ( \mathcal T_{P,L}f_L +\mathcal T_{P,R}f_R)
\nonumber\\
&&\int d\epsilon (\mathcal T_{P,L}+\mathcal T_{P,R})\bar f_P =
\int d\epsilon (\mathcal T_{L,P}f_R +\mathcal T_{R,P}f_L).
\label{eq:fpp}
\eea
We further assume that  $f_P(\phi)$, $f_P(-\phi)$ and $\bar f_P(\phi)$ have
a Fermi-Dirac form. We obtain the respective- unique \cite{unique}
chemical potentials by solving these equations (separately)
numerically-iteratively using
the Newton-Raphson method \cite{NR}.

{\it (ii) Results for $\phi=2\pi k$.} When the magnetic phase is
given by multiples of $2\pi$, we note that $\mathcal T_{\nu,\xi}=
\mathcal T_{\xi,\nu}$, particularly $\mathcal T_{L,P}=\mathcal
T_{P,L}=\frac{\gamma_L}{\gamma_R}\mathcal T_{P,R}$, with
$\gamma_{\nu}$ as the hybridization strength of the $\nu$ reservoir
to the system, see discussion around Eq. (\ref{eq:gamma}). Using the
voltage
probe condition (\ref{eq:fpp}) we find that
%
%\bea
%\int
% (\mathcal T_{P,L}+\mathcal T_{P,R})(f_P+\bar f_P)d\epsilon
%= \int (\mathcal T_{P,L}+\mathcal T_{P,R})(f_L+f_R)d\epsilon,
%\nonumber \eea
%
%leading to
$\int \mathcal T_{P,\mu}(f_P+\bar f_P)d\epsilon
= \int \mathcal T_{P,\mu}(f_L+f_R)d\epsilon$, $\mu=L,R$,
providing $\mathcal R=0$ in Eq. (\ref{eq:R}).

{\it (iii) Spatially symmetric setups.} If the
junction is left-right symmetric, the mirror symmetry $\mathcal
T_{P,L}(\phi)=\mathcal T_{P,R}(-\phi)$ applies. This translates to
the relation
%
%\bea
$\mathcal T_{P,L}(\phi)=\mathcal T_{R,P}(\phi)$.
%\label{eq:R3}
%\eea
%
We plug this result into Eq. (\ref{eq:fpp}) and note
that $\bar f_P(\phi)=f_P(-\phi)$.
The deviation from phase rigidity, Eq. (\ref{eq:DeltaI}), can also be
expressed as
%using the current flowing into the $R$ terminal,
%
\bea
\Delta I&=& \frac{1}{2}\int d\epsilon[(\mathcal T_{L,R}-\mathcal T_{R,L})f_L
\nonumber\\
&-&\mathcal T_{R,P}f_P(-\phi) +\mathcal T_{P,R}f_P(\phi)].
\label{eq:DeltaI2}
\eea
We now define $\Delta I$ by the average of Eqs.
(\ref{eq:DeltaI}) and (\ref{eq:DeltaI2}),
\bea \Delta I&=& \frac{1}{4}\int d\epsilon\Big[(\mathcal
T_{L,R}-\mathcal T_{R,L})(f_L+f_R)
\nonumber\\
&+&(\mathcal T_{L,P}-\mathcal T_{R,P})f_P(-\phi) +(\mathcal
T_{P,R}-\mathcal T_{P,L})f_P(\phi)\Big]. \nonumber \eea
Using Eq. (\ref{eq:R2}),
we note that $\mathcal T_{L,R}-\mathcal T_{R,L} =
\mathcal T_{P,L}-\mathcal T_{L,P}$. Furthermore, given that
$\mathcal T_{P,L}=\mathcal T_{R,P}$ in geometrically symmetric junctions
we get
\bea
&& \Delta I= \frac{1}{4}\int (\mathcal T_{P,L}-\mathcal
T_{P,R})(f_L+f_R-f_P-\bar f_P) d\epsilon
\nonumber\\
&&= \mathcal R(\phi)=- \mathcal R(-\phi)
\eea
%
%see Eq. (\ref{eq:R}).
Thus, in spatially symmetric systems, comprising inelastic interactions with an effective bias dependency, odd conductance terms acquire even symmetry with
respect to the magnetic field, $\mathcal D(\phi)=\mathcal D(-\phi)$,
as noted experimentally \cite{breakE2, breakE4},
%$G_{2k+1}(\phi)=G_{2k+1}(-\phi)$,
while even conductance terms, constructing $\mathcal R$, are odd
with respect to $\phi$. Next we show that these observations do not
generally hold when a spatial asymmetry is introduced, by coupling the
scattering centers unevenly to the leads.

%=========================================

%{\it (iv) Results for geometrically asymmetric junctions.}
{\it (iv) Double-dot interferometer.}
We perform numerical simulations for an AB device with a quantum dot
located at each arm of the interferometer. The Hamiltonian
includes the terms
\bea
H=H_S+\sum_{\nu=L,R,P}H_{\nu} +
\sum_{\nu=L,R} H_{S, \nu}
+H_{S,P},
\label{eq:H}
\eea
where the subsystem Hamiltonian includes two uncoupled electronic states,
and the three reservoirs (metals) comprise of a collection of non-interacting
electrons,
\bea H_S=\sum_{n=1,2}\epsilon_na_n^\dagger a_n,  \,\,\,\,\,
H_{\nu}=\sum_{j\in \nu}\epsilon_ja_j^{\dagger}a_j.
\eea
Here $a_{j}^{\dagger}$ ($a_{j}$)  are fermionic creation
(annihilation) operators of electrons with momentum $j$ and energy
$\epsilon_j$. $a_{n}^{\dagger}$ and $a_{n}$ are the respective
operators for the dots. The subsystem-bath coupling terms are given by
\beq H_{S,L}+H_{S,R}= \sum_{n,l} v_{n,l} a_{n}^{\dagger}a_{l}
e^{i\phi_{n}^{L}}+\sum_{n,r}v_{n,r}a_{r}^{\dagger}a_{n}
e^{i\phi_{n}^{R}}+h.c.,
\nonumber
\eeq
and we assume that only dot '1' is coupled to the probe
%
%\bea
$H_{S,P}= \sum_{p} v_{p} a_{1}^{\dagger}a_{p}+ h.c$.
%\eea
%
%Here $n={1,2}$ denotes dots '1' and '2',
Here $v_{n,\nu}$ is the coupling
strength of dot $n$ to the $\nu$ bath. Below we assume
that this parameter does not depend on the dot index.
$\phi_{n}^{L}$ and $\phi_{n}^{R}$ are the AB phase factors, acquired by electron
waves in a magnetic field perpendicular to the device plane. These
phases are constrained to satisfy
%
%\beq
$\phi_{1}^{L}-\phi_{2}^{L}+\phi_{1}^{R}-\phi_{2}^{R}=\phi$.
%\eeq
%
In what follows we adopt the gauge
$\phi_{1}^{L}-\phi_{2}^{L}=\phi_{1}^{R}-\phi_{2}^{R}=\phi/2$. We
voltage-bias the system, $\Delta \mu\equiv\mu_L-\mu_R\geq0$,
%with $\mu_{L,R}$ as the chemical potential of the metals.
in a symmetric manner, $\mu_L=-\mu_R$. However, the
dots energies may be placed away from the so called ``symmetric
point" at which $\mu_L-\epsilon_{n}=\epsilon_{n}-\mu_R$ using a gate
voltage.
Our model does not include interacting particles, thus its
steady-state characteristics can be readily obtained using the
nonequilibrium Green's function approach \cite{Wingreen}. Transient
effects were recently explored in \cite{Salil1, Ora-time}.
%We have recently adopted this method for the study of
%occupation behavior in degenerate and symmetric AB interferometers
%\cite{Salil2}.
%Since all the relevant derivations are included in
%earlier works, particularly in Ref. \ref{Salil2}, we only write here
%the main results.
%It can be also shown that
%the charge current through the $\nu$ terminal obeys a Landauer type expression \cite{Comm}
%%
%\bea
%I_{\nu}(\phi)&=&\frac{1}{2\pi} \int_{-\infty}^{\infty}d\epsilon \Big[\sum_{\xi \neq \nu}
%\mathcal T_{\nu,\xi}(\epsilon,\phi)
%f_{\nu}(\epsilon)
%\nonumber\\
%&-&\sum_{\xi \neq \nu}\mathcal T_{\xi,\nu}(\epsilon,\phi)f_{\xi}(\epsilon) \Big].
%\label{eq:curr}
%\eea
%
%As was mentioned above, the probe distribution, the result
%of the probe condition, is generally phase dependent
%beyond the symmmetric point.
%This property will be exemplified and discussed in Sec. XXX.
%
In terms of the Green's function, the transmission coefficient is
defined as $\mathcal T_{\nu,\xi}={\rm
Tr}[\Gamma^{\nu}G^+\Gamma^{\xi}G^-]$; the trace is performed over
the states of the subsystem. Given our Hamiltonian, the
matrix $G^{+}$ ($G^-=[G^+]^{\dagger}$) takes the form \cite{Salil2}
\bea G^{+}=\left[  \begin{array}{cc}
 \epsilon-\epsilon_{1}+\frac{i(\gamma_L+\gamma_R+\gamma_P)}{2} & \frac{i\gamma_L}{2}e^{i\phi/2}
+\frac{i\gamma_R}{2}e^{-i\phi/2}\\
\frac{i\gamma_L}{2}e^{-i\phi/2}+\frac{i\gamma_R}{2}e^{i\phi/2}
&  \epsilon-\epsilon_{2}+\frac{i(\gamma_L+\gamma_R)}{2}\\
 \end{array}\right]^{-1},
\nonumber \eea
with the hybridization matrices
\bea &&\Gamma^{L}=\gamma_L\left[ \begin{array}{cc}
              1 & e^{i\phi/2}\\
              e^{-i\phi/2} &1\\
\end{array}\right],\,\,\,\,
\Gamma^{R}=\gamma_R \left[ \begin{array}{cc}
              1 & e^{-i\phi/2}\\
              e^{i\phi/2} &1\\
\end{array}\right]
\nonumber\\
&& \Gamma^{P}=\gamma_P\left[ \begin{array}{cc}
              1 & 0\\
              0 & 0\\
\end{array}\right]
\label{eq:gamma}
\eea
The coupling energy between the dots and leads is given by
$\gamma_{\nu}(\epsilon)=2\pi\sum_{j\in\nu}|v_j|^2\delta(\epsilon-\epsilon_j)$.
In our calculations we take $\gamma_{\nu}$ as energy independent
parameters. It is interesting to note that the transmission
functions are not necessarily even in $\phi$, even when Onsager
symmetry is maintained. We exemplify this by considering %for simplicity
a geometrically symmetric system with
$\epsilon_d\equiv\epsilon_1=\epsilon_2$ and $\frac{\gamma}{2}\equiv
\gamma_L=\gamma_R$. The transmission functions reduce to
\bea
&&\mathcal T_{L,R}(\epsilon,\phi)= \mathcal T_{R,L}(\epsilon,-\phi)
\nonumber\\
&&=\frac{4\gamma^2}{\Delta(\epsilon,\phi)}
\Big[4(\epsilon-\epsilon_d)^2\cos^2{\frac{\phi}{2}}+
\frac{\gamma_{P}^2}{4}
+\gamma_{P}
(\epsilon_d-\epsilon)\sin{\phi}\Big],
\nonumber\\
&&\mathcal T_{L,P}(\epsilon,\phi) = \mathcal
T_{P,L}(\epsilon,-\phi)= \mathcal T_{P,R}(\epsilon,\phi)
\nonumber\\
&&
=\frac{4\gamma\gamma_{P}}{\Delta(\epsilon,\phi)}
\Big[2(\epsilon-\epsilon_d)^2+\frac{\gamma^2}{2}\sin^2\frac{\phi}{2}
+\gamma(\epsilon-\epsilon_d)\sin\phi\Big].
\nonumber
\label{eq:T2}
\eea
%
%$\mathcal T_{R,P}(\epsilon,\phi) = \mathcal T_{P,L}(\epsilon,\phi)$.
%
The denominator $\Delta(\epsilon,\phi)$ is an even function of phase.
%
%\bea
%\Delta(\epsilon,\phi)&=&16 \Bigg\{ \left[ (\epsilon-\epsilon_d)^2 -\frac{\gamma^2}{4} \sin^2 \frac{\phi}{2}
%-\frac{\gamma\gamma_P}{4}
%\right]^2
%\nonumber\\
%&+& \left[ (\epsilon-\epsilon_d)^2\left(\gamma+\frac{\gamma_P}{2}\right)^2  \right]\Bigg\}
%\eea
%
Due to the presence of the probe, these functions combine odd and
even magnetic field terms. When the probe dephases the system, we
substitute these expressions into Eq. (\ref{eq:dephP}) and resolve
the dephasing ($D$) probe distribution \cite{Salil2,comm2}
%
%\bea
%&&
$ f_P^D(\epsilon,\phi)= \frac{f_L(\epsilon)+f_R(\epsilon)}{2}
%\nonumber\\
%&&
+ \frac{\gamma(\epsilon-\epsilon_d)\sin \phi
}   {4\Big[(\epsilon-\epsilon_d)^2 + \omega_0^2\Big]}
[f_L(\epsilon)-f_R(\epsilon)]$ %\label{eq:fp} \eea
with $\omega_0=\frac{\gamma}{2}\sin \frac{\phi}{2}$.
The nonequilibrium
term in this distribution is {\it odd} in the magnetic
flux. Similarly,
when a voltage probe ($V$) is implemented, analytic results can be obtained
in the linear response regime upon solving Eq. (\ref{eq:fpp}),
%We assume the probe distribution
%to follow a Fermi function, and obtain the
%chemical potential
%
\bea
\mu_P^V(\phi)= \frac{\Delta \mu}{2}  \sin \phi
\frac{\int d\epsilon \frac{\partial f_a}{\partial \epsilon}
\frac{\gamma(\epsilon-\epsilon_d)}{\Delta(\epsilon,\phi)}}
{\int d\epsilon \frac{\partial f_a}{\partial \epsilon}
\frac{2(\epsilon-\epsilon_d)^2 +\frac{1}{2}\gamma^2\sin^2 \frac{\phi}{2}}{\Delta(\epsilon,\phi)}}.
\label{eq:muPG}
\eea
Here $f_a$ stands for the equilibrium (zero bias) Fermi-Dirac
function. This chemical potential is an {\it odd} function of the
magnetic flux, though phase rigidity is maintained in the linear
response regime.
%It is identically zero at the symmetric point.

%===================

%===================
\begin{figure}[tb]
\hspace{2mm} %\vspace{8mm}
{\hbox{\epsfxsize=60mm \epsffile{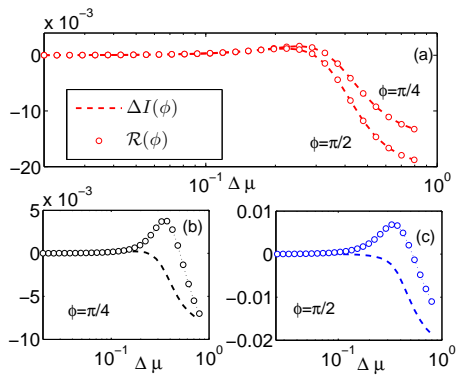}}}
\caption{
%Symmetry relations for nonequilibrium transport.
(a) In spatially symmetric junctions $\gamma_L=\gamma_R=0.05$,
%even conductance terms are non-even in the magnetic flux,
$\mathcal R=\Delta I$. (b)-(c) When spatial asymmetry is introduced,
$\gamma_L=0.05$ and $\gamma_R=0.2$, $\mathcal R\neq\Delta I$. We
adopt a voltage probe mimicking inelastic effects with
$\gamma_P=0.1$, $\epsilon_1=\epsilon_2=0.15$ and inverse temperature
$1/T_{\nu}=50$. The bands are taken broad and flat with a linear
dispersion relation. In all plots $\mathcal R$ is represented by
$\circ$ and $\Delta I$ by dashed lines.} \label{Fig1}
\end{figure}

\begin{figure}[t]
\hspace{1mm} %\vspace{2mm}
{\hbox{\epsfxsize=60mm \epsffile{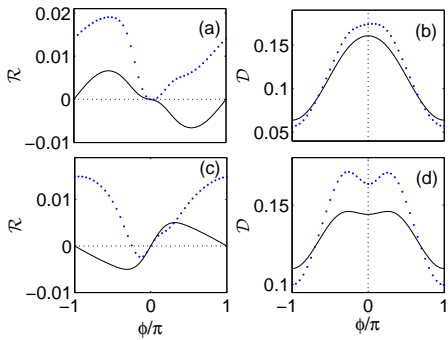}}}
\caption{
%Symmetry relations for nonequilibrium transport.
(a) Even ($\mathcal R$)
and (b) odd  ($\mathcal D$)
conductance terms in spatially symmetric junctions,
$\gamma_L=\gamma_R=0.05$ (full) and with an asymmetry
$\gamma_L=0.05\neq\gamma_R=0.2$ (dotted)
for $\epsilon_1=\epsilon_2=0.15$.
(c)-(d) Same as above, only with nondegenerate levels
$\epsilon_1=0.1$ and $\epsilon_2=0.2$.
Parameters are the same as in Fig. \ref{Fig1}, $\Delta\mu=0.4$.
Light dotted lines present the symmetry lines.
} \label{Fig2}
\end{figure}

%=================================

We adopt the model (\ref{eq:H}) and implement inelastic effects with a
voltage probe, by solving the probe condition (\ref{eq:fpp})
numerically-iteratively \cite{NR} to obtain $\mu_P$
beyond linear response. We have verified that when
convergence is reached the probe current is negligible,
$|I_P/I_L|<10^{-12}$.
%Exact numerical results are displayed next.
In Fig. \ref{Fig1} we show that $\mathcal R=\Delta I$
if spatial symmetry is maintained, and that this relation is
violated when $\gamma_L\neq \gamma_R$. Note that phase rigidity,
$\Delta I=0$, persists in the linear response regime
%$\Delta \mu \lesssim\epsilon_n$
\cite{Buttiker-4}. % in all cases.
In Fig. \ref{Fig2} we further extract the sum of odd conductance terms
%using $\mathcal D= I_L-\mathcal R$,
and confirm that it strictly satisfies
 $\mathcal D(\phi)= \mathcal D(-\phi)$  in
spatially symmetric situations. Interestingly,
while we noted that the symmetry of $\mathcal R$ is feasibly broken with
small spatial asymmetry, the symmetry of $\mathcal D$ is more robust
and deviations are very
small even when $\gamma_R\gg \gamma_L$, in support of
experimental observations \cite{breakE2}. Our
conclusions are intact when an ``up-down" asymmetry is
implemented in the form
$\epsilon_1\neq\epsilon_2$, see Fig. \ref{Fig2}(c)-(d) \cite{comm3}

%=================================

{\it Summary.} We presented an analytical and numerical study of
nonlinear transport properties of AB rings susceptible to elastic
dephasing and inelastic effects by adopting the B\"uttiker probe
method. We proved that  $\mathcal D(\phi)=\mathcal D(-\phi)$ and
$\mathcal R(\phi)=-\mathcal R(-\phi)$ for spatially symmetric junctions, though many-body inelastic
 effects, introduced via the probe, are nonlinear in the applied bias.
We also demonstrated the strong and weak breakdown of these symmetries when the
junction has a left-right asymmetry, in the presence of inelastic effects.
It is of interest to  verify these results
adopting a microscopic model with many-body interactions, modeling a
quantum point contact \cite{QPC1,QPC2} or an equilibrated phonon bath,
%exchanging energy with the junction's electronic degrees of freedom,
% that can exchange energy with electronic degrees of freedom,
by extending recent works \cite{Hod, Ora} to the nonlinear regime.

The work of SB has been supported by NSERC and through the ERA award of DS.

%---------------------------------------------------------------------

%-----------------------------


\begin{thebibliography}{32}
\bibitem{comm}
The transmission function $\mathcal
T_{\nu,\xi}(-\phi)$ is presented here in its complementary form,
$\mathcal T_{\xi,\nu}(\phi)$.
\bibitem{comm2}
Eq. (41) of Ref. \cite{Salil2} suffers from a typo, missing
correct transmission functions; results
below Eq. (41) are correct.
Eq. (\ref{eq:IP}) here introduces the correct form, besides a
$2\pi$ factor that is absorbed in the defintiion $h=1$.
\bibitem{comm3}
It is interesting to note that at the symmetric point, $\epsilon_1=\epsilon_2=0$
the symmetries $\mathcal D(\phi)=\mathcal D(-\phi)$ and  $\mathcal R(\phi)=-\mathcal R(-\phi)$  are obeyed, even when the left-right symmetry is broken.
\end{thebibliography}
\end{document}